\documentclass[conference]{IEEEtran}
\IEEEoverridecommandlockouts
\pdfoutput=1
\usepackage{cite}
\usepackage{amsmath,amssymb,amsfonts}
\usepackage{algorithmic}
\usepackage{graphicx}
\usepackage{textcomp}
\usepackage{xcolor}
\def\BibTeX{{\rm B\kern-.05em{\sc i\kern-.025em b}\kern-.08em
    T\kern-.1667em\lower.7ex\hbox{E}\kern-.125emX}}
\begin{document}

\title{Resting state-fMRI approach towards understanding impairments in mTLE\\}

\author{\IEEEauthorblockN{Hritik Bansal\thanks{* Both authors contributed equally to this manuscript} \footnotemark*}
\IEEEauthorblockA{\textit{Electrical Engineering Department} \\
\textit{Indian Institute of Technology Delhi}\\
India \\
ee1160071@iitd.ac.in}
\and
\IEEEauthorblockN{Nishad Singhi \footnotemark*}
\IEEEauthorblockA{\textit{Electrical Engineering Department} \\
\textit{Indian Institute of Technology}\\
India \\
ee1160107@iitd.ac.in}
}

\maketitle

\begin{abstract}
Mesial temporal lobe epilepsy (mTLE) is the most common form of epilepsy. While it is characeterised by an epileptogenic focus in the mesial temporal lobe, it is increasingly understood as a network disorder. Hence, understanding the nature of impairments on a network level is essential for its diagnosis and treatment. In this work, we review recent works that apply resting state functional MRI to provide key insights into the impairments to the functional architecture in mTLE. We discuss changes on both regional and global scales. Finally, we describe how Machine Learning can be applied to rs-fMRI data to extract resting state networks specific to mTLE and for automated diagnosis of this disease. 
\end{abstract}

\begin{IEEEkeywords}
Independent components, fMRI. resting state, Default mode network, perceptual networks, Machine learning (ML), graph theory, feature selection
\end{IEEEkeywords}

\section{Introduction}
Epilepsy in general is associated with heterogenity in clinical history, neuropathological condition, neurophysiological features, electroencephalographic and neuroimaging findings. Mesial temporal lobe epilepsy (mTLE) is epilepsy of temporal origin which is defined by characteristic hippocampal sclerosis (HS) as its pathological substrate. mTLE is considered to be surgically treatable condition, and efforts are being taken to understand its characteristics precisely \cite{b28}.

Non-invasive imaging technique such as functional magnetic resonance imaging (fMRI) use stimulus driven paradigms to delve into the intracacies of brain function. Recently, there has been an enormous increase in interest in the application of this technique at rest, which termed as resting state fMRI (rs-fMRI).

rs-fMRI may provide greater insight into the functional connectivity in the neural systems, rather than what is being activated by a particular task. It also helps in eliminating the noise associated with atypical strategies related to task design and the sensitivity and specificity of various metrics. \cite{b29}, \cite{b30}.

A number of methods are available to analyze the rs-fMRI data like seed-based analysis, Independent component analysis (ICA) and graph based approach. Out of these, seed-based analysis requires apriori selection of Regions of Interests (RoIs) based on predetermined templates. This method requires thresholding to identify voxels that are significantly correlated with the RoI. In this study, we will be looking at ICA specifically because of its various advantages, as highlighted in Section \ref{sec: ica}. Graph based methods have revealed that the brain exhibits small-world topology, which allows nodes (individual RoIs) to have low number of connections while still being connected to all other nodes with a short distance \cite{b29}. We touch upon this in our study in section \ref{sec: graph}.

With the advent of machine learning (ML) and computational resources, it has been possible to learn complex functions that fit the data. Studies have used hand-engineered features in some cases, and in the other let the model derive those features automatically from the raw data using more sophisticated techniques like deep learning. In this study \ref{ml_fmri}, \ref{sec: ml_graph_features}, we highlight how machine learning techniques have been exploited recently for better understanding and discovery of epileptic networks which can be missed by sheer correlation-based statistical methods.
studies 
\section{Independent Components from fMRI data}
\label{sec: ica}
\begin{figure}[h]
\centerline{\includegraphics{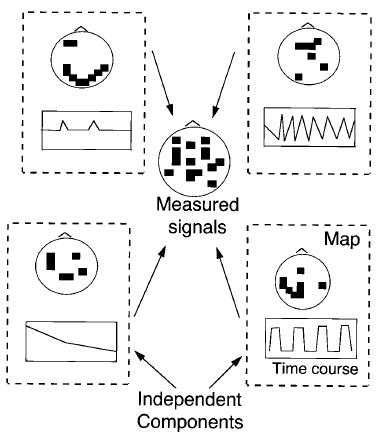}}
\caption{fMRI data decomposed into spatial independent components. \cite{b1}. First column of the mixing matrix (M) gives the time course modulation of the first IC, second column gives the time course modulation of the second IC and so on.}
\label{ica}
\end{figure}

Analytical techniques like time-frequency analysis, Analysis of variance (ANOVA) and principal component analysis (PCA) have several drawbacks pertaining to extraction of useful information from changes in fMRI data (includes blood oxygen level-dependent (BOLD) signal). Generally, cognitive/perceptual task related fMRI signals are typically small ($<10\%$), suggesting other time-varying phenomena must be contributing to the bulk of the measured signal \cite{b1}.\\
In time-frequency analysis, the time signals associated with each voxel are converted into fourier (frequency) domain. Such techniques assume that the task performed by the subject causes distinguishable changes in the frequency  spectrum of the data. Some techniques also assume the source signals of the data to be periodic, which is infact a strong assumption to make. ANOVA \cite{b4},\cite{b5} makes certain assumptions on the prior distribution of the data (e.g. Gaussian), time series associated with each voxel being independent and variances between repeated measurements are equal. However, such techniques do not extract the intrinsic structure of the data, and do not capture transient task related changes.\\
PCA \cite{b3} has long been considered as a way to identify the tendency of signals at each pair of voxels to covary. It finds the spatial patterns (eigenvectors) which capture the greatest variance of the data. However, when task related signal is very small in comparison to the bulk, getting eigenvectors (images) associated with greatest variance in the data might not be useful. There is a tendency to miss overall associations in the brain networks using this technique when multiple voxels get activated simultaneously \cite{b1}.  \\
Independent Component Analysis (ICA) \cite{b1},\cite{b6} technique posits that separate processes associated with multifocal brain areas may be represented by multiple spatially-independent components. Each independent component has a single time signal and spatial map associated with it (Fig. \ref{ica}). We presume that high-order correlations between the voxels values in the pairs of components is zero. ICA is posed as solving:

\begin{equation}
    X = MC
\end{equation}
\begin{equation}
    C = WX
\end{equation}

where $C \in R^{n \times K}$ is a matrix of independent components, rows of which consists of n independent components with K being the number of voxels. $M \in R^{T \times n}$ is the mixing matrix, which controls the contribution of each independent component to the observed signal $X \in R^{T \times K}$. T is the number of time datapoints in the observed signal. W is the unmixing signal, which helps in retrieving the underlining components C. We estimate \textit{W} Using an iterative unsupervised learning algorithm \cite{b6} based on mutual information principles and without making an prior assumptions on the activation's spatiotemporal extent.\\ 
Rather than considering the observed signal as a linear combination of independent components, non-linear ICA \cite{b8} poses the problem as:
\begin{equation}
    \mathbf{x(t)} = f(\mathbf{s(t)})
\end{equation}
where \textbf{x(t)} is an n-dimensional data point at time t, and f is smooth and invertible mixing function, and \textbf{s(t)} is an n-dimensional vector of independent components $\mathbf{s_{i}(t)}$. Even the time series $\mathbf{s_{i}}$ are assumed to be mutually independent. \cite{b2}, \cite{b7} show that non-linear ICA of the fMRI data can express functional traits associated with resting state, cognitive and perceptual task in the human brain. 

\section{Resting State Networks}
\label{sec: rsn}
\begin{figure}[h]
\centerline{\includegraphics[scale=0.6]{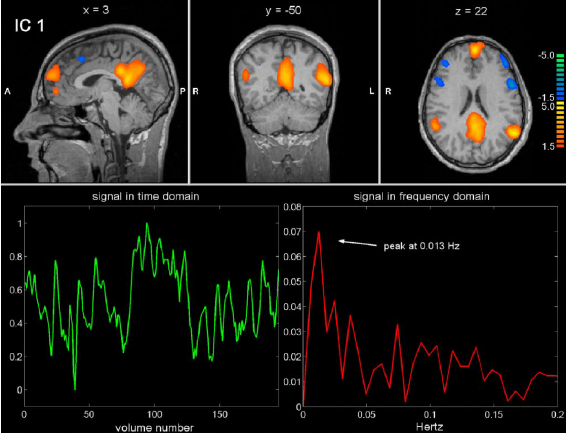}}
\caption{Example of 1 IC seperated from a subject using ICA \cite{b15}. The figure shows the sagittal, coronal and axial spatial map of the IC. Orange colored brain areas show positive correlation with the IC waveform, and the blue ones show negative correlation.}
\label{electrophysio1}
\end{figure}

Positron emission tomography (PET) and functional Magnetic Resonance Imaging (fMRI) studies \cite{b10} have provided evidence of transient changes (\textit{activations}) induced by cognitive and perceptual tasks. Several studies \cite{b11}, \cite{b14} have suggested that fMRI images obtained using blood oxygen level dependent (BOLD) contrast show signal fluctuations at rest (awake or anesthetized brain). \\
The regions displaying coherent low frequency (0.01-0.1 Hz) fluctuations constitute a \textit{resting state network} (RSN). The understanding of the neurophysiological basis of these RSNs helps in identifying the functional role of spontaneous activity. \cite{b15} identify resting state patterns from the BOLD signals by using Independent Component Analysis (ICA) (Fig. \ref{electrophysio1}). 

Across subjects, \cite{b15} identified six RSNs, as exemplified in the Table \ref{RSN}. 

\begin{table}[h]
\label{tab:rsnets}
\centering
\begin{tabular}{l | r | r}
RSN \# & Regions & Task\\
\hline
RSN 1 & Default Mode Network (DMN) & Cognitive\\
RSN 2 &   Dorsal Attention Network & Cognitive\\
RSN 3 &  Occipital cortex & Perceptual\\
RSN 4 &  Temporal cortex & Perceptual\\
RSN 5  &  Sensory Motor & Perceptual\\
RSN 6  &  Self-referential mental activity & Cognitive\\

\end{tabular}
\caption{\label{RSN} Resting state networks and their corresponding regions and task that they are associated with.}
\end{table}

Several studies \cite{b15}, \cite{b21} have shown the correlation between the resting state fMRI and electroencephalography (EEG) power data collected simultaneously from the subjects. EEG-fMRI and rs-fMRI have frequently shown networks related to epileptic transients, characterization of which becomes important in understanding the neurological differences in the controls vs epilepsy patients. 

\section{Structural and Functional abnormalities in DMN}
\label{sec: dmn}
The focal point in mTLE is usually in the mesial temporal region (hence the name), usually hippocampal sclerosis (HS). But, mTLE is increasingly being understood as a network disorder, meaning that its effects can extend to regions far beyond the mesial temporal lobe. Previous studies have shown that epileptic activity can extend from the temporal lobe to other regions of the `Default mode network' \cite{b9}, which is a network of regions that are \textit{functionally connected} to each other and are usually active during wakeful rest. It has been shown that functional and structural connectivity overlap in the case of DMN and that structural organization is directly related to the functional synchronization of DMN \cite{b12} \cite{b13}. Hence, \cite{b16} studied how structural and functional connectivity changes in the DMN.

Functional connectivity was measured using Pearson correlation (in fMRI) and structural connectivity was measured using connection density (number of connections per unit surface) and path length (average length of all connecting fibres). As compared to controls, patients showed a decrease in both connection density and temporal correlation between posterior cingulate cortex (PCC)/precuneus (PCUN) and both mesial temporal sclerosis (mTLs). These results suggest that the sturctural connecticity between these two areas is degraded and along with this, decreased functional connectivity may cause DMN abnormalities in patients suffering from mTLE. Patients showed a decrease in the maximum fractional anisotropy value between PCC/PCUN and medial prefrontal cortex (mPFC). Apart from this, no significant difference was found between these two regions. Interestingly, mean temporal correlation was found to be significantly correlated with connection density between PCC/PCUN and mTLs in both groups, further suggesting that functional connectivity and structural connectivity between these areas are related. Broadly, these results suggest that while HS is the epileptogenic focus, it may cause structural degradation of connections in DMN (like the connections between PCC/PCUN and mTLs), which contributes to functional decline in patients.

\begin{figure}[h]
\centerline{\includegraphics[scale = 0.6]{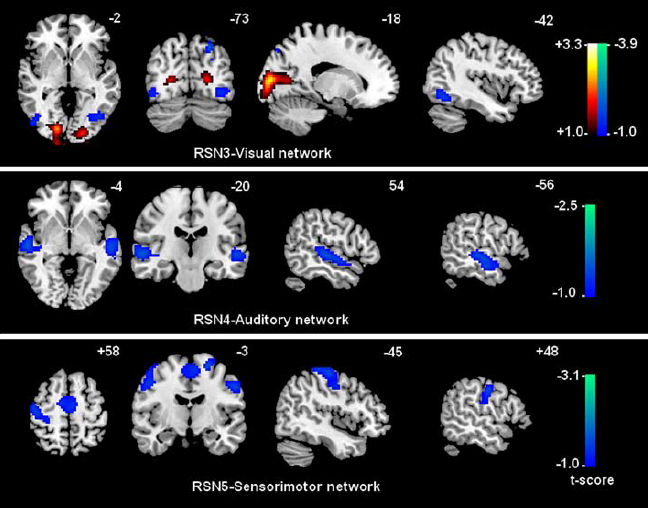}}
\caption{Functional connectivity differences between the controls and mTLE patients \cite{b23}. Blue indicates decreased connectivity within a region for patients when compared with the controls; Red indicates the increased connectivity.}
\label{perceptual_net}
\end{figure}

\section{rsfMRI and Perceptual networks in mTLE}

mTLE has shown to adversely affect the cognitive functioning such as memory, largely associated with DMN \ref{sec: dmn}, and language \cite{b22}. Studies have also shown impairments in the attention as well as perceptual networks \cite{b23}, \cite{b24}. Perceptual networks are linked to the functionality of the visual, auditory and sensorimotor systems (Table \ref{RSN}). ICA based analysis of BOLD-fMRI data in resting state has provided significant insights into neurophysiological mechanisms underlying functional impairments in mTLE. The results \cite{b23} obtained from this analysis of RSN agreed with earlier findings \cite{b25}, \cite{b26}. By performing correlation based statistical analysis of the RSNs corresponding to perceptual networks of the mTLE patients and controls, we can assess the functional connectivity differences between the two Fig. \ref{perceptual_net}. It is important to note as per Fig. \ref{perceptual_net} the increased functional connectivity within primary cortex might be linked to the plasticity behavourial changes occuring in the mTLE patients in response to impairments in other parts like bilateral MT+ areas. \\
In \cite{b23}, the authors show that mean z-values within the region of interests (RoIs) from Fig. \ref{perceptual_net} are negatively correlated with a clinical parameter: epilepsy duration, but uncorrelated to seizure frequency.

\section{Identification of rs-fMRI networks in TLE using ML}
\label{ml_fmri}
Resting-state networks were introduced in Section \ref{sec: rsn}. While ICA (see Section \ref{sec: ica}) can extract a large number of RS networks, research in fMRI typically focuses on a few well-studied resting state networks (see Table \ref{RSN}) because their spatial structure is consistent with the understanding of brain functions developed using task-based fMRI. However, powerful machine-learning techniques have made it possible to study the remaining components derived by ICA. \cite{b17} analysed rs-fMRI data of 42 patients and 90 controls with the aim of identifying ICs that could be indicative of TLE.

After extracting 88 ICs using ICA, top 10 ICs were identified with the help of an elastic-net based feature selection method. These networks included frontal, temporal, perisylvian, cingulate, posterior-quadrant, thalamic, and cerebellar regions. The intensity of these ICs were correlated with clinical variabled and hippocampal volumes and it was found that many of these components were significantly correlated with duration of epilepsy, number of anti-epileptic drugs, duration of epilepsy, etc. Two ICs showed significant correlation with the affected hippocampal volume and not with the unaffected one.

The significance of this study is clear from the fact that the networks identified using ML correlated strongly with several clinical variables, indicating that they are specific to this disease. Moreover, the strengths of these ICs was only related to the affected hippocampus and not the unaffected one, which again supports the epilepsy-specific nature of these ICs. Also, IC strength is negatively correlated with hippocampal volume which implies that hippocampal volume is lost as the disease condition worsens, which is consistent with previous findings \cite{b19}. Finally, the method achieves accuracy that is comparable to previous studies using different techniques.

\section{Graph theory based approaches}
\label{sec: graph}
\subsection{Altered functional connectivity and small world}
Several studies have shown that functional connectivity within DMN is altered in patients suffering from mTLE. However, most of these studies focus on local changes in connectivity, analysing changes in individual connections. \cite{b20} analysed rs-fMRI data of 18 patients and 27 controls using graph theoretical measures that provide a more global picture of differences between patients and healthy subjects.

90 RoIs were defined in the brain and their correlation matrix was computed. Each region served as a node in a graph and an edge existed between the two nodes if their correlation was above a certain threshold. Several graph theory measures were computed and compared between the two groups. Significant differences were found between the two groups for several topological variables, suggesting a macroscopic reorganization in mTLE. It was found that connectivity decreased in the frontal, parietal, and occipital lobes, which are some of the key areas in DMN. It is known that both functional and structural connectivity in the DMN are altered in patients (See Section \ref{sec: dmn}), hence these findings are consistent with the literature. Significant negative correlation between rIFGoper and lIFGtri with the epilepsy duration was found, providing further evidence that changes in functional connectivity affect observable clinical variables. Along with decreased functional connectivity in the DMN, an increase in functional connectivity was observed in other areas of the brain, which suggests that epilepsy may arise from an imbalance of connectivity in regions of the brain. 

Several important regions showed decreased values of degree in patients, which suggests that their connectivity to other regions of the brain was affected, which may inhibit the flow of information. Particularly, PCC/PCUN, which is a crucial area in the DMN (see Section \ref{sec: dmn}), showed a decrease in the value of degree. There were significant differences in the n-to-1 connectivity values of several regions across the groups, suggesting altered functional connectivity. 

The clustering coefficients were smaller in patients, suggesting that the graphs were sparser. Shorter path lengths were found in patients which indicates faster and more efficient interactions between regions. Finally, the small-world properties were altered in patients in a way that their graphs were closer to random graphs. These findings suggest that graph theory measures can be used as markers for this disease. 

\subsection{Using ML and graph-based features to detect lateralization in mTLE}
\label{sec: ml_graph_features}
One of the most important aspects of preseurgical testing in mTLE is successful lateralization of the affected hemisphere. Usually, this is done by manual assessment of structural MRI images, which leads to accurate classification in 70-85\% cases. As we have previously seen, functional MRI provides rich information about changes in mTLE. \cite{b27} employed graph-based features as used in the previous study and machine learning to develop an automated lateralization pipeline (henceforth referred to as CADFIG).

\begin{figure}[h]
\centerline{\includegraphics[scale = 0.275]{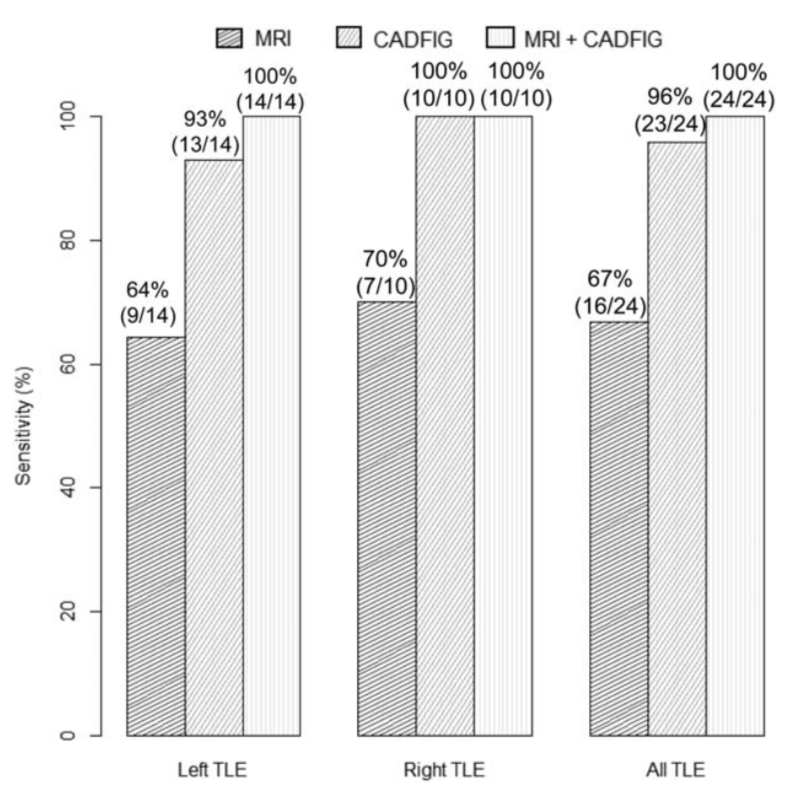}}
\caption{Classification results of CADFIG, Manual assessment (denoted by MRI) and combined approaches \cite{b27}. }
\label{fig:ML_graph}
\end{figure}

CADFIG could accurately lateralize 95.8\% of the patients, which was better than manual assessment (66.7\% accuracy). Moreover, CADFIG showed a 42.9\% increase in sensitivity for right TLE and 44.4\% increase for left TLE (see Figure \ref{fig:ML_graph}). Clearly, CADFIG is superior than manual assessment, which is treated as the `gold standard' in presurgical evaluation of lateralization. When the two methods were used together, all patients were classified correctly, which means that a multimodal approach may work best. Since functional connectivity could correctly lateralize most of the patients, it can be said that right and left mTLE differ as far as functional connectome is concerned.

\section{Conclusions and future directions}
In this survey we reviewed a powerful data driven technique called `Independent Component Analysis' (ICA, Section \ref{sec: ica}) and described how it can be applied to extract spatial networks of functionally connected regions called `resting state networks' from fMRI data (Section \ref{sec: rsn}). we focus on two categories of these networks -- the `default mode network' which pertains to wakeful rest, self awareness; and `perceptual networks', which involve regions important for visual, auditory, and sensorimotor processing. Using both structural MRI (DTI) and rs-fMRI, it was shown that degradation of connections can spread from HS to other regions of DMN, which may subsequently affect the functional connectivity leading to impairments in patients. Similarly, significant alterations were observed in the perceptual resting state networks.

These studies on resting state networks focus on the well-studied networks (see Table \ref{tab:rsnets}). However, using powerful ML techniques, other independent components derived from fMRI data can be studied. It was showed that using an elastic-net based feature selection algorithm, `epilepsy specific' ICs could be identified in patients and their strengths correlated with clinical variables. Additionally, most studies focus on regional or local changes in connectivity. Using graph-theory based measures, it was showed that mTLE induces global topological changes in functional connectivity. Further, these measures were used as features in a ML algorithm and provided significant improvement over manual assessment of structural MRI to detect lateralization in patients.

Recent improvements in ML have shown promise in advancing the literature on applications of rs-fMRI to mTLE. Traditional methods of statistical analysis require manually fixed thresholds whereas ML techniques are free from this. Moreover, ML techniques also provide insights about the information contained in specific features (for example, by using feature selection techniques). ML also makes it possible to develop tools for automated diagnosis of mTLE, which cannot be done using statistical analyses alone. ML also makes it possible to capture complicated, non-linear patterns in the data. Hence, we suggest that this direction should be further pursued. Specifically, it would be worthwhile to apply non-linear ICA techniques to capture richer spatio-temporal patterns of connectivity in the brain. 

In addition to this, there are evidences of neurological differences across genders and taking them into account can even improve the diagnostic ability of ML systems \cite{b31} which makes it important to investigate if there are any systematic differences in the functional connectivity in male and female patients of mTLE.

\bibliographystyle{plain}

\end{document}